\documentclass[review]{elsarticle}
\pdfoutput=1

\usepackage{amssymb,amsmath,latexsym,mathrsfs}
\usepackage{graphicx} 
\usepackage{bm}
\usepackage[normalem]{ulem}
\usepackage{hyperref}
\usepackage{comment}
\hypersetup{
    colorlinks=true,       
    }   
 \usepackage{xspace}
\usepackage{soul,xcolor}

\newcommand{\cosmomc}{\texttt{CosmoMC}\xspace}

\newcommand\be{\begin{equation}}
\newcommand\ee{\end{equation}}
\newcommand\bea{\begin{eqnarray}}
\newcommand\eea{\end{eqnarray}}
\newcommand\mpl{M_{\rm Pl}}
\newcommand\fnl{f_{\text{NL}}}

\newcommand{\pt}{\,\,\,\,\,\,}

\newcommand\ie{{\it i.e.}~}
\newcommand\eg{{\it e.g.,}~}

\begin{document}

\setstcolor{red}

\begin{frontmatter}

\title{Primordial power spectrum features in phenomenological descriptions of inflation}

\author[infnTo,unito]{Stefano Gariazzo}
\author[ific]{Olga Mena}
\ead{omena@ific.uv.es}
\author[ific]{H\'ector Ram\'irez}
\author[quito]{Lotfi Boubekeur}

\address[infnTo]{Department of Physics, University of Torino,
Via P. Giuria 1, I--10125 Torino, Italy}
\address[unito]{INFN, Sezione di Torino,
Via P. Giuria 1, I--10125 Torino, Italy}
\address[ific]{Instituto de F\'isica Corpuscular (IFIC), CSIC-Universitat de Valencia,\\ 
Apartado de Correos 22085,  E-46071, Spain}
\address[quito]{Universidad San Francisco de Quito USFQ, Colegio de Ciencias e Ingenier\'ias El Polit\'ecnico,\\
campus Cumbay\'a, calle Diego de Robles y V\'ia Interoce\'anica, Quito EC170157, Ecuador.}

\begin{abstract}
We extend an alternative, phenomenological approach to inflation by means of an equation of state and a sound speed, both of them functions of the number of $e$-folds and four phenomenological parameters. This approach captures a number of possible inflationary models, including those with non-canonical kinetic terms or scale-dependent non-gaussianities. 
We perform Markov Chain Monte Carlo analyses using the latest cosmological publicly available measurements, which include Cosmic Microwave Background (CMB) data from the Planck satellite. Within this parametrization, we discard scale invariance with a significance of about $10\sigma$, and the running of the spectral index is constrained as $\alpha_s=-0.60\,^{+0.08}_{-0.10}\times 10^{-3}$ ($68\%$~CL errors). The limit on the tensor-to-scalar ratio is $r<0.005$ at $95\%$~CL from CMB data alone. We find no significant evidence for this alternative parameterization with present cosmological observations. The maximum amplitude of the equilateral non-gaussianity that we obtain, $|f^{\text{equil}}_{\text{NL}}|< 1$, is much smaller than the current Planck mission errors, strengthening the case for future high-redshift, all-sky surveys, which could reach the required accuracy on equilateral non-gaussianities. 

\end{abstract}

\end{frontmatter}


\section{Introduction} 

Inflation is the leading and most attractive theory, with observational success, capable of describing the initial conditions of the universe while solving the main problems of the standard Big Bang Cosmology~\cite{Guth:1980zm,Linde:1981mu,Albrecht:1982wi}. Usually inflation is described via the dynamics of a single new scalar degree of freedom, the {\sl inflaton}, coupled to Einstein Gravity and slowly-rolling down a potential. The validity of the proposed inflationary potential relies on its predictions for the standard inflationary observables: the tensor-to-scalar ratio $r$, characterizing the amplitude of the gravitational waves produced during inflation, the scalar spectral index $n_s$, measuring the scale dependence of the power spectrum $P_\zeta(k)$, its running $\alpha_s$ and possibly, the running of the running $\beta_s$. However there is a plethora of theoretical models belonging to this type of scenario, \ie of inflationary potentials, that could give predictions of the inflationary observables that are in good agreement with current Cosmic Microwave Background (CMB) measurements~\cite{Ade:2015lrj}. According to these observations, structures grow from Gaussian and adiabatic primordial perturbations. 

However, another probe of the mechanism of the inflationary physics comes from the study of non-Gaussian components of the primordial fluctuations~\cite{Bartolo:2004if}. These contributions are characterized by the three-point correlation function of the primordial curvature perturbations $\zeta$ or its Fourier transform, the bispectrum $B_\zeta(k)$. It is well known that a detectable large amount of non-gaussianities would rule out the standard single-field slow-roll scenarios~\cite{Acquaviva:2002ud,Maldacena:2002vr}, leading to the study of exotic inflationary models or even theories with different dynamics for the generation of primordial perturbations. The amount of non-gaussianities is characterized by the observable $f_{\text{NL}}$ defined as $\zeta({\bf x})=\zeta_\text{g}({\bf x})+f_\text{NL}^{\text{local}}(\zeta_{\text{g}}({\bf x})^2-\langle\zeta_\text{g}({\bf x})^2\rangle)$, where $\zeta_{\text{g}}$ is the primordial Gaussian curvature perturbation~\cite{Komatsu:2001rj,Komatsu:2003iq}. Recent measurements from Planck CMB polarization data have set the limits $\fnl^{\text{local}}=0.8\pm5.0$, $\fnl^{\text{equil}}=-4\pm43$ and $\fnl^{\text{ortho}}=-26\pm21$ with $68\%$~CL errors~\cite{Ade:2015ava}.

In general, there are inflationary models in which the value of the sound speed of the primordial curvature perturbation, $c_s$, can be different from that of the speed of light~\footnote{In single-field slow-roll inflation $c_s=1$.}. These models are characterized by allowing non-canonical kinetic terms in the Lagrangian (see, \eg\cite{Chen:2006nt} and references therein). Theoretically, it is possible to derive a limit in the sound speed as a function of the tensor-to-scalar-ratio, provided that $c_s$ is constant~\cite{Baumann:2014cja}. Models in which not only the sound speed is non-standard (\textit{i.e.}\ $c_s\neq 1$) but also varies with time, such as in Dirac-Born-Infeld (DBI) inflation~\cite{Chen:2005fe,Bean:2008na,Miranda:2012rm}, lead to an amplitude $\fnl$ of the primordial bispectrum which is scale-dependent~\cite{LoVerde:2007ri,Sefusatti:2009xu}. A varying sound speed, $c_s=c_s(\tau)$, during the inflaton evolution can imprint features in both the matter power spectrum and bispectrum ($P_\zeta(k)$ and $B_\zeta(k)$, respectively)~\cite{Park:2012rh,Achucarro:2012fd}, see Ref.~\cite{Chluba:2015bqa} for an extensive review~\footnote{Features in the primordial power spectrum may also arise in inflationary models with a sharp step/feature in the inflaton potential~\cite{Adams:2001vc,Hunt:2004vt} (see also Refs.~\cite{Peiris:2003ff,Covi:2006ci,Jain:2008dw,Jain:2009pm,Mortonson:2009qv,Hazra:2010ve,Benetti:2011rp,Adshead:2011bw,Adshead:2011jq,Bartolo:2013exa,Miranda:2013wxa,Miranda:2014wga,Miranda:2015cea,Cadavid:2015iya,Benetti:2016tvm,Chen:2016vvw}), or in axion monodromy scenarios~\cite{Silverstein:2008sg,McAllister:2008hb,Flauger:2009ab,Huang:2012mr,Easther:2013kla,Flauger:2014ana,Motohashi:2015hpa}, (see also the recent work of Ref.~\cite{Hazra:2016fkm}).}. These signatures can be constrained using cosmological data and, therefore, studying them also helps as a discriminator of the inflationary mechanisms.

In this work we adopt the phenomenological description of inflation from~\cite{Mukhanov:2013tua} in which both the equation of state and the sound speed are parameterized as a function of the number of $e$-folds $N$. The usual inflationary parameters, \textit{i.e.}\ tensor-to-scalar ratio $r$, the scalar spectral index $n_s$ and its running $\alpha_s$, are derived quantities and will also depend on $N$. As we shall illustrate in the following, this model generates features in the primordial power spectrum. The deviations from the standard $P_\zeta(k)$, due to the variation of the sound speed, and the generated amplitude of the bispectrum, $\fnl$, will be exploited to constrain this phenomenological approach to inflation.

The structure of the paper is as follows. Section~\ref{sec:Muk} describes the parametrization used in this study. In Sec.~\ref{sec:Ach} we use the available tools to compute the features in the primordial power spectrum $P_\zeta(k)$ and the scale-dependent non-gaussianities arising on models with non-constant sound speed, applying them to our particular case. Section~\ref{sec:data} contains the description of the method and of the cosmological data sets. In Sec.~\ref{sec:Res}, we present our results, including the derived limits on the standard inflationary parameters. Finally, we draw our conclusions in Sec.~\ref{sec:conclusions}.

\section{Phenomenological approach to inflation}
\label{sec:Muk}

An alternative approach to describe the inflationary paradigm can be provided by a phenomenological parametrization based on a hydrodynamical picture, through an equations of state~\cite{Mukhanov:2013tua,Barranco:2014ira,Boubekeur:2014xva}. During inflation, the equation of state is $p\simeq-\rho\simeq-3H^2\mpl^2$, while $p\ll\rho$ towards its ending. Here $p$ and $\rho$ are the pressure and the energy density respectively, $H$ is the Hubble parameter and $\mpl$ the reduced Planck mass~\footnote{As usual, $M_{\rm pl}=1/\sqrt{8\pi G_N}\simeq 2.43\times 10^{18}$ GeV.}. With this scenario in mind, one can thus write a parametrization of the equation of state in terms of the number of $e$-folds left to the end of inflation, $|\text{d}N|\equiv H\text{d}t$, as

\be
\frac{p}{\rho}+1=\frac{\beta}{(N+1)^{\alpha}}\pt.
\label{eos}
\ee
Here the parameters $\alpha$ and $\beta$ are both positive and of order unity. As shown in Refs.~\cite{Mukhanov:2013tua,Barranco:2014ira} the parametrization above captures different inflationary models which vastly differ in their observational signatures. This hydrodynamical characterization of the inflationary scenario allows as well for a non-standard, time-varying $c_s$. The sound speed is parameterized via~\cite{Mukhanov:2013tua}

\begin{equation}
c_s=\frac{\gamma}{(N+1)^{\delta}}~,
\label{cc}
\end{equation}
where $\delta\geq 0$, because the sound speed is assumed to grow towards the end of inflation, and $\gamma$ is an arbitrary positive number. The expressions for the derived quantities, as the tensor-to-scalar ratio, the primordial spectral index and its running read as~\cite{Barranco:2014ira}:
\be
\begin{aligned}
r&=\frac{24\,\beta\,\gamma}{(N_\star+1)^{\alpha+\delta}}\pt,\\ \\
n_s&=1-\frac{3\beta}{(N_\star+1)^\alpha}-\frac{\alpha+\delta}{N_\star+1}\pt,\\ \\
\alpha_s&=-\frac{3\alpha\beta}{\left(1+N_\star\right)^{\alpha+1}}-\frac{\alpha+\delta}{\left(N_\star+1\right)^2}\pt,
\label{derived}
\end{aligned}
\ee
where $N_\star$ indicates the number of remaining $e$-folds at horizon crossing. Notice that both the running $\alpha_s$ and the tilt, $n_s-1$, are always negative.


\section{Features in $P_\zeta(k)$ and $f_\text{NL}$}
\label{sec:Ach}

It has been shown that a consequence of the inflationary models with a varying sound speed of the primordial perturbations is the presence of features in the primordial power spectrum $P_\zeta(k)$ and in the primordial bispectrum $B_\zeta(k)$~\cite{Achucarro:2012fd,Achucarro:2013cva,Achucarro:2014msa,Palma:2014hra,Mooij:2015cxa,Chluba:2015bqa}. The deviations of the primordial power spectrum from the standard case can be studied by isolating the contributions due to a non-standard $c_s$ as $P_\zeta(k)=P_0(k)+\Delta P_\zeta(k)$. Here $P_0(k)=H^2/(8\pi^2\epsilon\mpl^2)$ is the usual featureless primordial power spectrum. The corrections to the primordial power spectrum generated by the sound speed variations through time are given by~\cite{Achucarro:2012fd}
\be
\frac{\Delta P_\zeta}{P_0}(k)=k\int^0_{-\infty}u(\tau)\text{sin}(2k\tau)d\tau\pt,
\label{int}
\ee
where $u(\tau)\equiv1-c_s^{-2}(\tau)$ and $\tau$ is the conformal time. Eq.~(\ref{int}) is valid if the reduction in the sound speed is small, \ie{}\ in the $|u(\tau)|\ll 1$ regime~\cite{Achucarro:2012fd}. 

Using $\text{d}N=-\text{d}\tau/\tau$, which is valid for a de-Sitter space-time with constant expansion rate, we can write Eq.~(\ref{cc}) in terms of the conformal time $\tau$, and therefore the corrections to the primordial power spectrum can be computed as
\be
\begin{aligned}
\frac{\Delta P_\zeta}{P_0}(k)=k\int^{\tau_0e^{N_{\rm{e}}}}_{\tau_0 e^{N_{\rm{i}}}}\left\{1-\gamma^{-2}\left[1+\text{ln}\left(\frac{\tau}{\tau_0}\right)\right]^{2\delta}\right\}\text{sin}(2k\tau)\text{d}\tau~,
\label{eq:deltap}
\end{aligned}
\ee
where $N_{\rm{e}}$ and $N_{\rm{i}}$ refer to the end and the beginning of the inflationary period, which, in this parametrization (see Eq.~(\ref{cc})), are identified with $N=0$ and $N\simeq 60$, respectively. In the following, we will fix the number of inflationary $e$-folds to $60$. 

\begin{figure*}[t]
\begin{tabular}{c c}
\includegraphics[width=.49\textwidth]{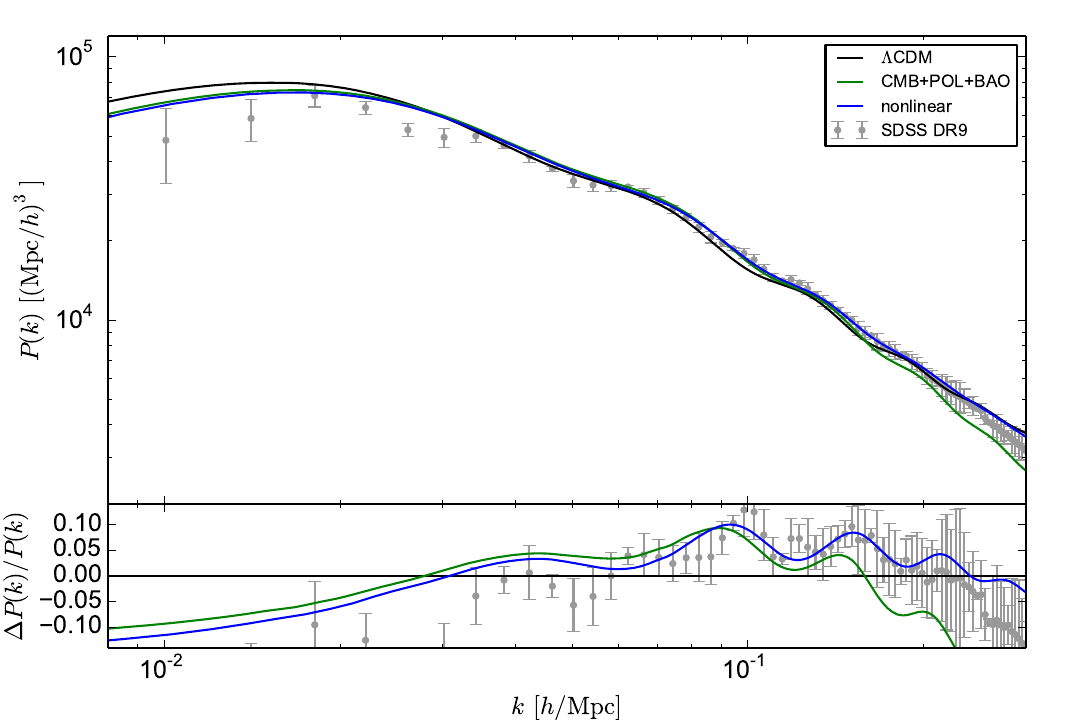} & \includegraphics[width=.49\textwidth]{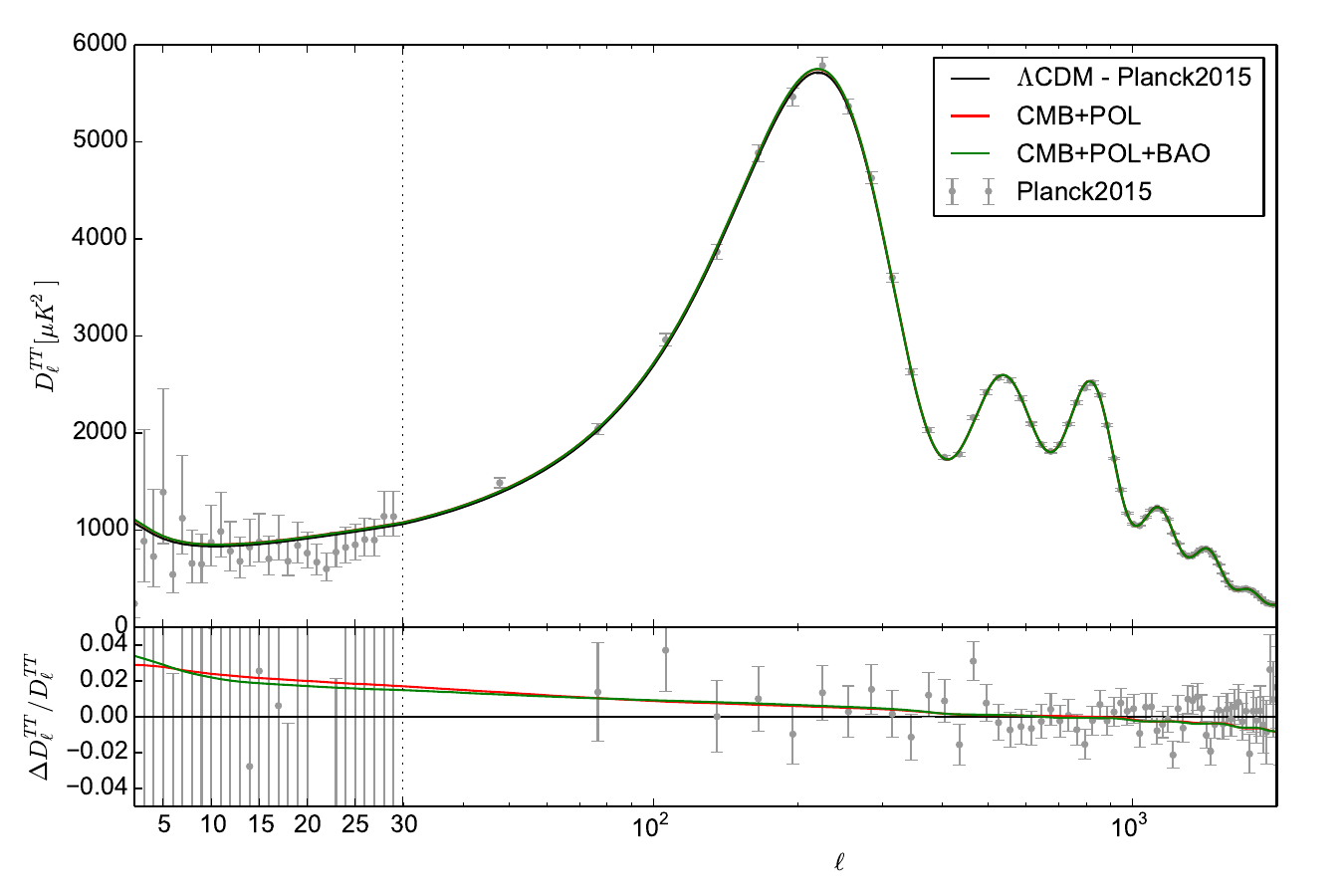} \\
\end{tabular}
\caption{\textit{Left panel, top:} Galaxy power spectrum for the best-fit in standard cosmologies (those with $c_s=1$, depicted by the black line) versus the results for the parametrization explored here, see text for details. The non-linear galaxy power spectrum corresponds to the prediction from the Coyote emulator of Kwan et al.~(2015)~\cite{Kwan:2013jva}. Data points are the clustering measurements from the BOSS Data Release
9 (DR9) CMASS sample~\cite{Ahn:2012fh}. \textit{Left panel, bottom:} Residuals with respect to the standard parametrization with $c_s=1$. \textit{Right panel, top:} Theoretical predictions for the temperature anisotropies (TT) in the standard best-fit $\Lambda$CDM cosmology (in black) and those obtained in the time-varying sound speed scenario (in red and green), together with the Planck 2015 TT data. \textit{Right panel, bottom:} Relative difference between the standard and non-standard sound speed schemes.}
\label{fig:power}
\end{figure*}

Figure~\ref{fig:power}, top left panel, shows the galaxy power spectrum $P(k)$ at a redshift $z=0.57$, which corresponds to the mean redshift of the DR9 CMASS sample of galaxies~\cite{Ahn:2012fh}. These power spectrum measurements will be exploited in the next section in their Baryon Acoustic Oscillation (BAO) form to set constraints on the phenomenological inflationary approach studied here. Together with these measurements we show, in the top panel, the galaxy power spectrum for the best-fit $\Lambda$CDM parameters for the standard inflationary parametrization with $c_s=1$~\cite{Ade:2015xua}, together with the galaxy power spectrum for a time-varying $c_s(\tau)$ scenario, fixing $\gamma=1$ and $\delta=0.032$~\footnote{This value of $\delta$ corresponds to the upper prior considered here for this parameter, see the following section.}. The non-linear galaxy power spectrum in the canonical ($c_s=1$) $\Lambda$CDM scheme corresponds to the prediction from the Coyote emulator of Kwan et al.~(2015)~\cite{Kwan:2013jva}. The bottom left panel of Fig.~\ref{fig:power} illustrates the deviations with respect to the pure-linear case with $c_s=1$. Notice, from the bottom panel, the oscillatory behaviour imprinted in the galaxy power spectrum, whose amplitude is governed by the parameter $\delta$. 

Figure \ref{fig:power} (top right panel) shows the Planck 2015 temperature anisotropies (TT) data~\cite{Adam:2015rua}, together with the theoretical predictions using the best-fit spectrum in a standard $\Lambda$CDM cosmology~\cite{Ade:2015xua} (\textit{i.e.}\ with $c_s=1$, see the black curve) and those obtained in the inflationary approach explored here (see Eq.~(\ref{cc})). For the non-canonical case, we depict two scenarios (see the red and green curves), associated to the $95\%$~CL upper bounds on the parameter $\delta$ governing the time-varying sound speed studied here, keeping $\gamma=1$, as the deviations of $c_s$ from $1$ can not be very large, otherwise Eq.~(\ref{int}) no longer applies. The remaining cosmological parameters have been set to their best-fit values. The bottom right panel of Fig.~\ref{fig:power} illustrates the relative difference between the best-fit model and the non-canonical cases, showing the difference among the models with a time-varying inflation sound speed.

The corrections to the bispectrum $B_\zeta(k)$ are parameterized in the same way, \ie $B_\zeta=B_0+\Delta B_\zeta$. Here, $B_0$ represents the contributions to the bispectrum when $c_s=1$ and $\Delta B_\zeta/B_0$ depends on $\Delta P_\zeta/P_0$ (see Ref.~\cite{Achucarro:2012fd} for details). The dimensionless shape function, $f_{\text{NL}}$, governing the amplitude of the primordial non-gaussianities, can be deduced for three configurations: equilateral, local and folded. Concretely, in the model we explore here, we shall focus on the equilateral type of non-gaussianity, which  typically arises in single field inflationary models with non-canonical kinetic terms and therefore with a  time-varying $c_s$. This type of non-gaussianities is given by~\cite{Achucarro:2012fd}:
\be
f^{\text{equil}}_{\text{NL}}= \frac5{54}\left[-7\frac{\Delta P_\zeta}{P_0}-3\frac{\text{d}}{\text{d\,ln\,}k}\left(\frac{\Delta P_\zeta}{P_0}\right)+\frac{\text{d}^2}{\text{d\,ln\,}k^2}\left(\frac{\Delta P_\zeta}{P_0}\right)\right]\bigg|_{k=\frac12(k_1+k_2+k_3)}\,,
\label{fnl}
\ee
with $k_2/k_1=k_3/k_1=1$.
The current limit on equilateral non-gaussianity from the Planck experiment including polarization data is $f^{\text{equil}}_{\text{NL}}=-4\pm 43$ ($68\%$~CL errors)~\cite{Ade:2015ava}. We shall illustrate in the next section that, for the values of the parameters within the ranges allowed by current cosmological data, the corresponding value of $f^{\text{equil}}_{\text{NL}}$ is much smaller than the present error bars on this quantity.

\section{Numerical analyses}
Throughout  this section, we shall present the constraints on the model described by Eqs.~(\ref{eos}) and (\ref{cc}) based on our numerical data analyses of current cosmological data. Both the methodology and the cosmological data sets exploited in this work are carefully detailed in what follows.

\subsection{Methodology and Cosmological data sets}
\label{sec:data}

To set constraints on the inflationary model described above, we used the Boltzmann solver CAMB~\cite{Lewis:1999bs} to compute the cosmological evolution. We fit the theoretical model to various cosmological datasets using the Markov Chain Monte Carlo code \cosmomc~\cite{Lewis:2002ah} to sample the parameter space.
Both codes were modified in order to compute the tensor to scalar ratio $r$, the scalar spectral index $n_s$ and the running $\alpha_s$ of the power spectrum of scalar perturbations as a function of $\alpha$, $\beta$ and $\delta$, fixing $\gamma\equiv 1$~\footnote{This choice is required to ensure the validity of Eq.~(\ref{int}).}, as in Eq.~(\ref{derived}), and to compute the deviations from the standard primordial power spectrum from Eq.~(\ref{eq:deltap}).
The integral in Eq.~(\ref{eq:deltap}) is computed using the \texttt{FILON} algorithm~\cite{Chase:1969a,Chase:1969b}, that is specifically designed to integrate rapidly oscillating functions. The power spectra of initial (scalar and tensor) perturbations are then described by the normalization $A_s$ and by the parameters $\alpha$, $\beta$, $\gamma\equiv 1$ and $\delta$ through the usual relations:
\be
\begin{aligned}
P_\zeta(k)&=\left(1+\Delta P_\zeta(k)\right)\, A_s\left(\frac{k}{k_0}\right)^{n_s-1+\alpha_s\ln(k/k_0)/2}\,,\\
P_t(k)&=r A_s\left(\frac{k}{k_0}\right)^{n_t+\alpha_t\ln(k/k_0)/2}\,,
\end{aligned}
\ee
where 
the $n_t$ and $\alpha_t$ are both obtained from the consistency relations $n_t=-r/8 (2-n_s-r/8)$ and
$\alpha_t=dn_t/d\ln k=r/8 (r/8+n_s-1)$,
and $k_0=0.05\,\mathrm{Mpc}^{-1}$ is the pivot scale.
The tensor-to-scalar ratio $r$ is referred to the pivot scale $k_0$.

The other parameters that we allow to vary are the baryon energy density $\Omega_{\rm b} h^2$, the cold dark matter energy density $\Omega_{\rm c} h^2$,
the size of the sound horizon at recombination $\theta_s$ and the optical depth to reionization $\tau$. For all these parameters we use the flat priors given in Tab.~\ref{tab:priors}. Notice that the priors chosen allow us to test a limited region in the sound speed parameter space, testing values of the sound speed very close to $1$. The reason for this limited region is due to the fact that the methodology followed here (see Eq.~(\ref{int})) requires the changes in the sound speed to be small. Larger departures of $c_s(\tau)$ from the canonical sound speed will require a calculation beyond the scope of this work. 

\begin{table}
\begin{center}
\begin{tabular}{c|c}
Parameter                    & Prior\\
\hline
$\Omega_{\rm b} h^2$ & $[0.005,0.1]$\\
$\Omega_{\rm c} h^2$ & $[0.001,0.99]$\\
$\theta_{\rm s}$     & $[0.5,10]$\\
$\tau$               & $[0.01,0.8]$\\
$\log[10^{10}A_{s}]$ & $[2.7,4]$\\
$\alpha$             & $[1.9,3]$\\
$\beta$              & $[0,1]$\\
$\log_{10}\delta$         & $[-4,-1.5]$\\
\end{tabular}
\end{center}
\caption{Priors for the parameters used in the \cosmomc analyses.}
\label{tab:priors}
\end{table}

We perform our analyses testing the theoretical predictions against the most recent CMB data from the 2015 release of the Planck collaboration~\cite{Aghanim:2015xee}. We separately consider the full temperature auto-correlation spectrum at all multipoles with the polarization spectra at low multipoles only (\textbf{Planck TT+lowP}) or with the inclusion of the polarization spectra at all multipoles (\textbf{Planck TT,TE,EE+lowP}).

To constrain the primordial tensor modes, we include in our analyses the joint results of the Bicep2/Keck and Planck collaborations~\cite{Ade:2015tva} (\textbf{BKP}). For sake of brevity, in the following we will indicate with \textbf{CMB} the combination \textbf{Planck TT+lowP+BKP} and with \textbf{CMB+POL} the combination \textbf{Planck TT,TE,EE+lowP+BKP}.

The last dataset that we consider involves the Baryon Acoustic Oscillation (\textbf{BAO}) measurements as obtained by several experiments. Namely, we exploit BAO measurements from the 6dFGS~\cite{Beutler:2011hx} at redshift $z=0.1$,  the SDSS Main Galaxy Sample (MGS)~\cite{Ross:2014qpa} at redshift $z_\mathrm{eff}=0.15$ and from the BOSS DR11 release at redshifts $z_\mathrm{eff}=0.32$ and $z_\mathrm{eff}=0.57$~\cite{Anderson:2013zyy}.

\subsection{Results}
\label{sec:Res}

The results arising from our numerical fits are summarized in Tab.~\ref{tab:results}, where we show the $95\%$~CL upper limits on the usual cosmological parameters ($\Omega_{\rm b} h^2$, $\Omega_{\rm c} h^2$, $\theta_{\rm s}$, $\tau$  and $\log[10^{10}A_{s}]$) as well as on the inflationary parameters describing the phenomenological model explored here: $\alpha$, $\beta$ and $\log_{10}(\delta)$, see Eqs.~(\ref{eos}) and (\ref{cc}). We present as well the corresponding limits on derived quantities, such as the Hubble constant $H_0$ and the clustering parameter $\sigma_8$, plus the constraints on the usual inflationary parameters $n_s$, $r$ and $\alpha_s$, obtained by making use of Eqs.~(\ref{derived}). We also show the derived limits on the sound speed during inflation, $c_s$. 

Table~\ref{tab:results} shows that, while the limits obtained on the $\alpha$ and $\beta$ parameters agree with previous findings in the literature~\cite{Barranco:2014ira}, the yet-unexplored $\delta$ parameter, directly related to the time-varying sound speed explored here, Eq.~(\ref{cc}), is not constrained by data. Values of $\gamma=1$ and $\delta=0$ lead to the canonical slow-roll scenario with a constant sound speed $c_s=1$. The prior choice assumed here for the parameter $\delta$ induces only mild departures in the sound speed with respect to this canonical scenario, leading to the allowed $3\sigma$ range $0.9<c_s<1$. The primordial power spectrum features induced by the time-dependent sound speed inflationary paradigm studied here are translated into non-negligible signatures in the photon temperature and polarization anisotropies as well as in the galaxy clustering power spectrum, as illustrated in the top panels of Fig.~\ref{fig:power}.

\begin{figure*}
\begin{center}
\begin{tabular}{c c c}
\includegraphics[width=.4\textwidth]{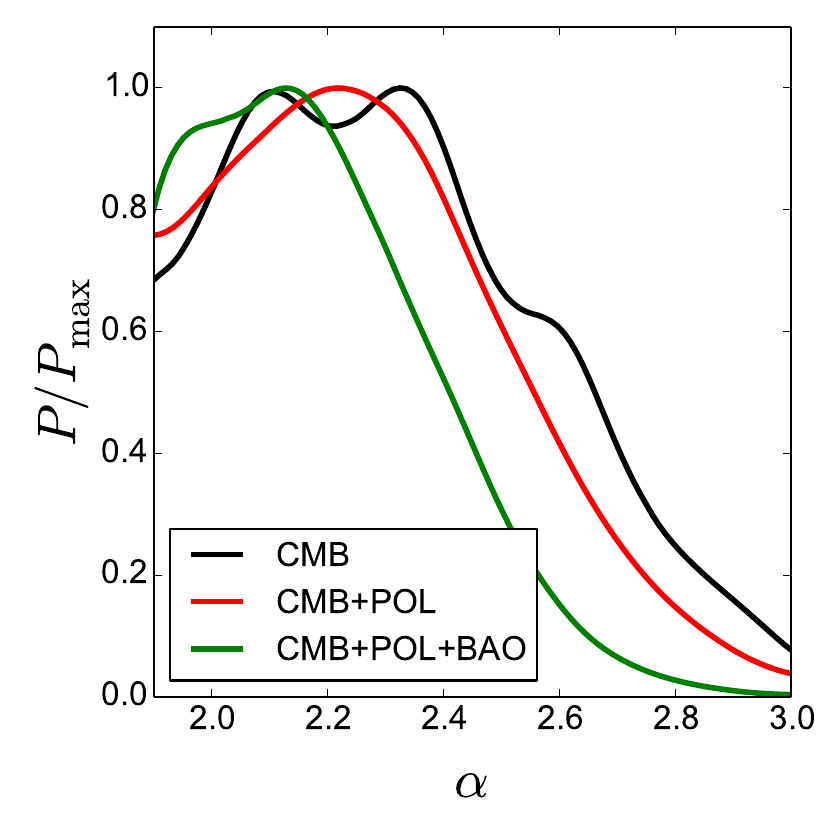} & \includegraphics[width=.4\textwidth]{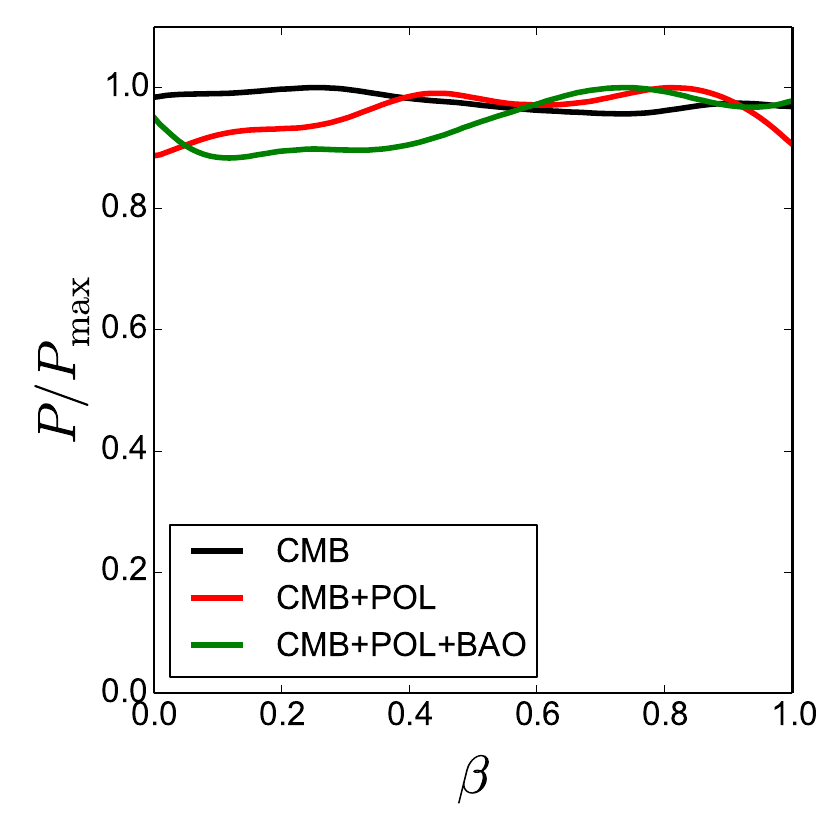} \\
\end{tabular}
\caption{One-dimensional posterior probability distributions for the $\alpha$ and $\beta$ parameters, involved in the phenomenological description of inflation studied here.}
\label{fig:1Dall}
\end{center}
\end{figure*}

The one-dimensional posterior probability distributions for $\alpha$ and $\beta$ parameters are illustrated in Fig.~\ref{fig:1Dall}. From these results we can notice that once that BAO information is included in the data analyses, \textit{(a)} the features in the power spectrum are required to be smaller; and \textit{(b)} due to the lower error on the $\alpha$ parameter found in this case, the deviation from a scale-invariant $n_s=1$ power spectrum will be more significant than with CMB data only, as we shall comment below. Notice as well that the parameter $\beta$ is unconstrained, as current data isolates the small-r region of Ref.~\cite{Barranco:2014ira}, which corresponds to $\alpha > 2$, with no preferred value of $\beta$, as the tensor-to-scalar ratio for this branch goes as $r\simeq 10^{-2} \beta$ (provided $\gamma=0$ and $\delta=1$). The value of the $\delta$ parameter is completely undetermined by present cosmological measurements.

\begin{figure}
\begin{center}
\begin{tabular}{c c}
\includegraphics[width=.4\textwidth]{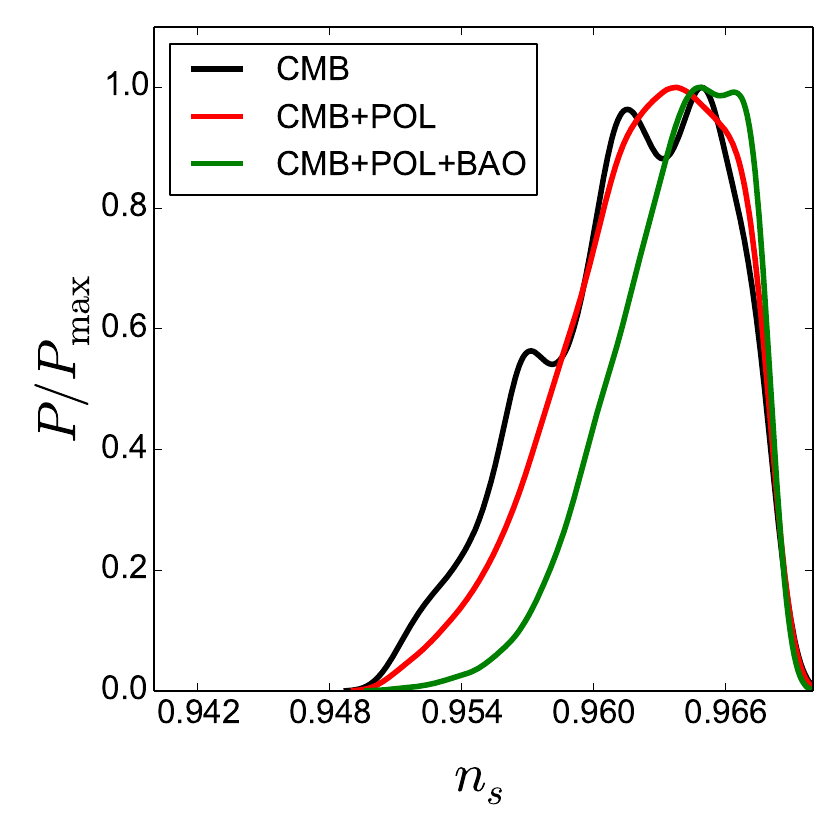}&
\includegraphics[width=.4\textwidth]{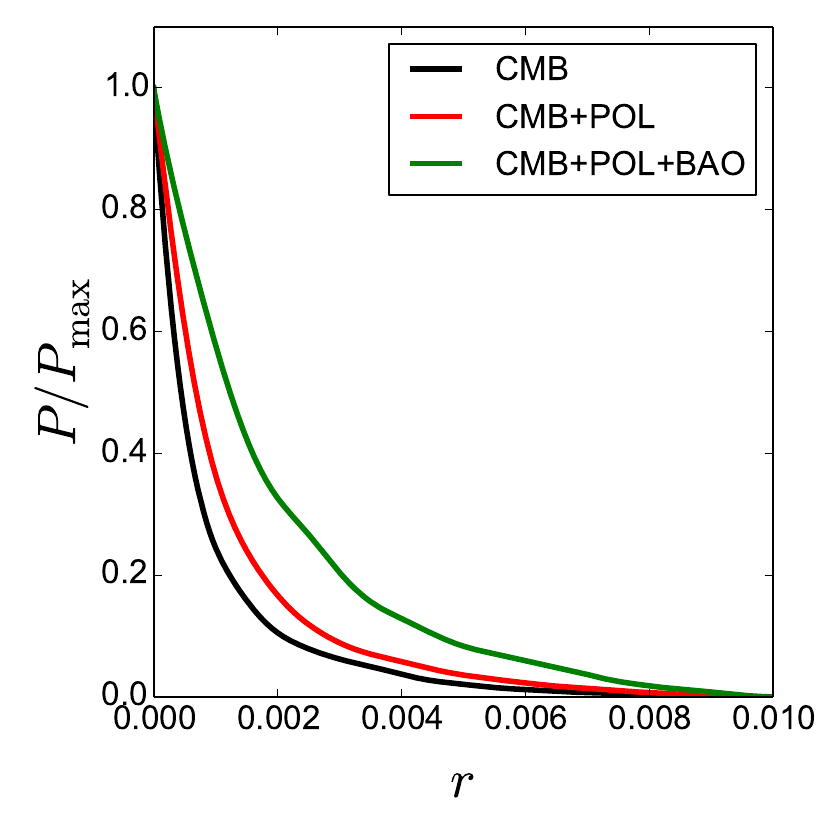}
\end{tabular}
\caption{One-dimensional posterior probability for the derived parameters $n_s$ and $r$. }
\label{fig:1Dallbis}
\end{center}
\end{figure}

Figure~\ref{fig:1Dallbis} illustrates the one-dimensional probability distributions for the standard inflationary parameters $n_s$ and $r$. The mean values and their associated $68\%$~CL errors we get for the scalar spectral index are $n_s= 0.962\pm 0.004$ and $n_s=0.963\pm 0.003$, for CMB or CMB+POL and CMB+POL+BAO respectively, rejecting scale invariance with a significance larger than $10\sigma$. As a comparison, when $n_s$ is considered a free parameter, the mean value and its $68\%$~CL error from CMB+POL are $n_s=0.965\pm 0.005$, discarding scale invariance at around $7\sigma$~\cite{Ade:2015xua}. 

\begin{figure*}
\begin{center}
\begin{tabular}{ccc}
\includegraphics[width=.32\textwidth]{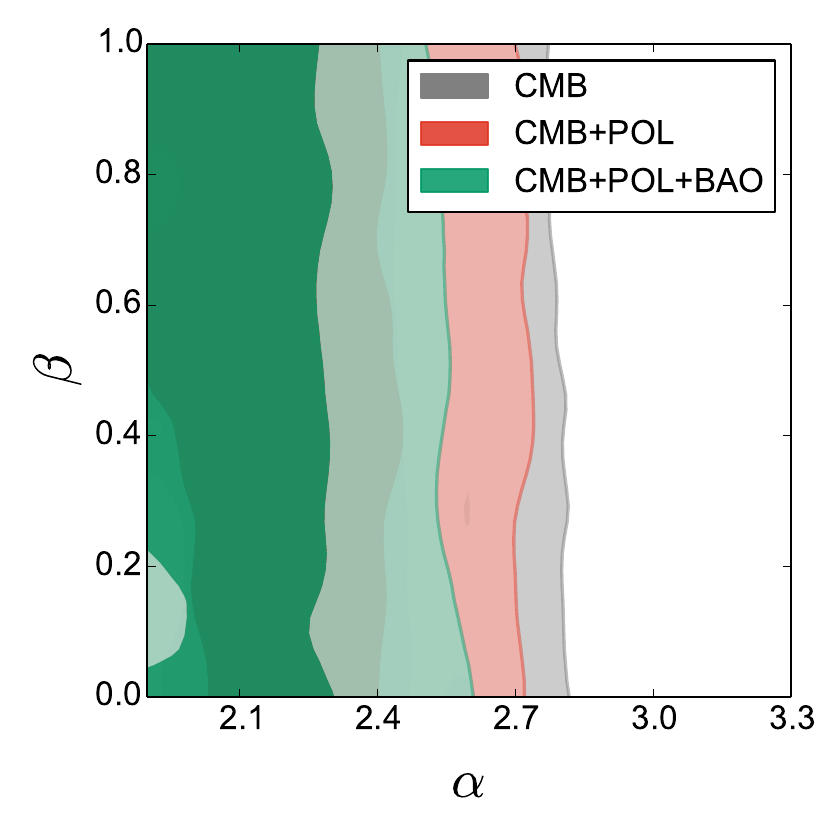}  &\includegraphics[width=.32\textwidth]{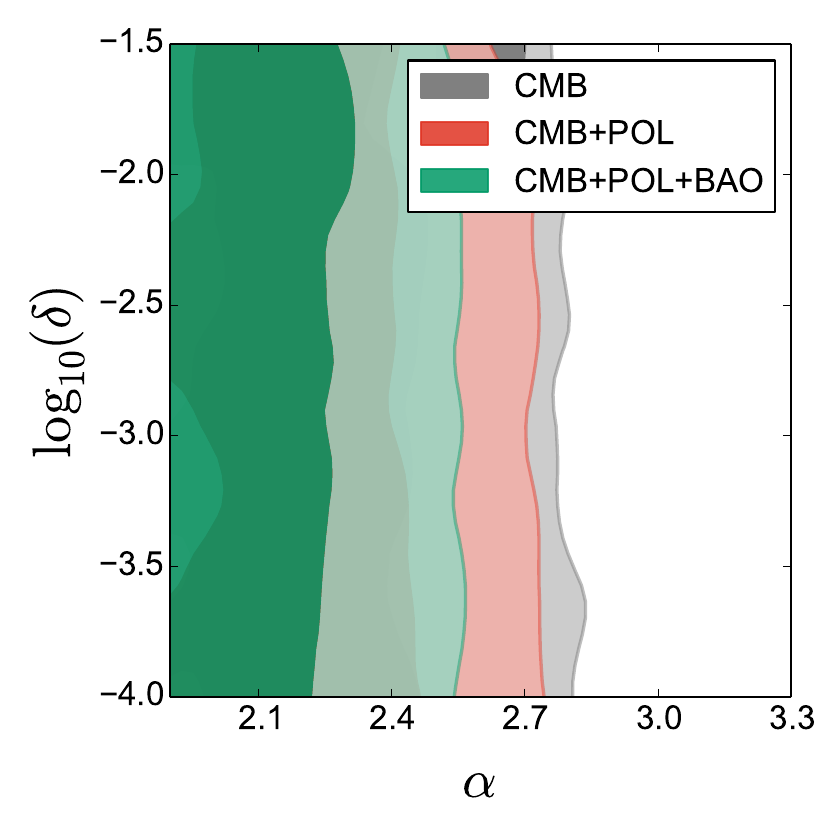} & \includegraphics[width=.32\textwidth]{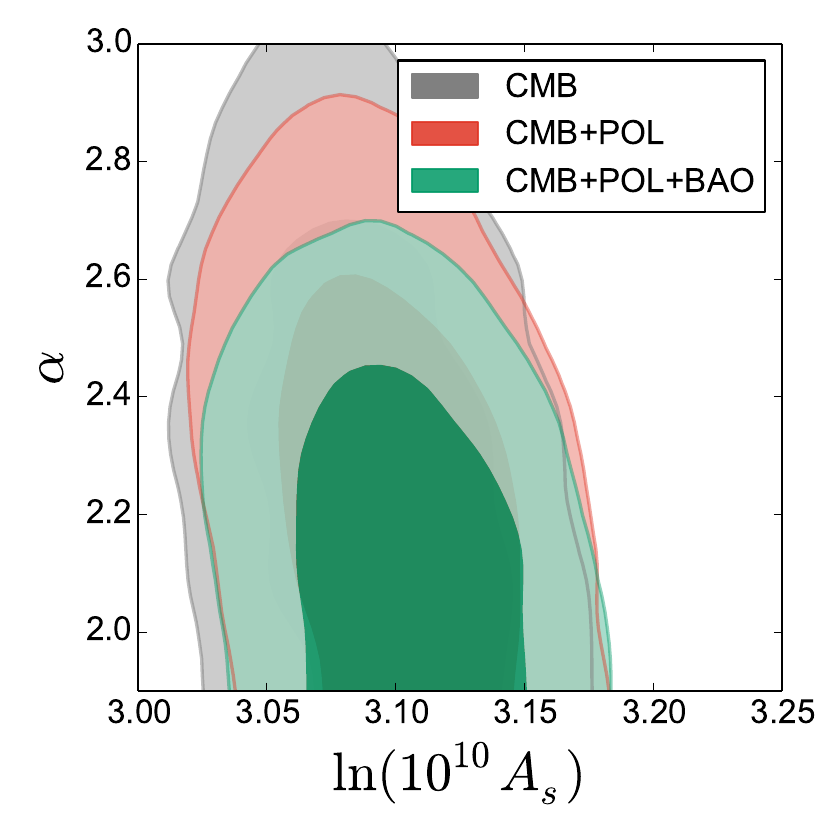}
\end{tabular}
\caption{From left to right: $68\%$ and $95\%$~CL allowed regions in the ($\alpha$, $\beta$), ($\alpha$, $\delta$) and ($\ln[10^{10}A_s]$, $\alpha$) planes.}
\label{fig:asnsbis}
\end{center}
\end{figure*}

\begin{figure}
\begin{center}
\begin{tabular}{c c}
\includegraphics[width=.4\textwidth]{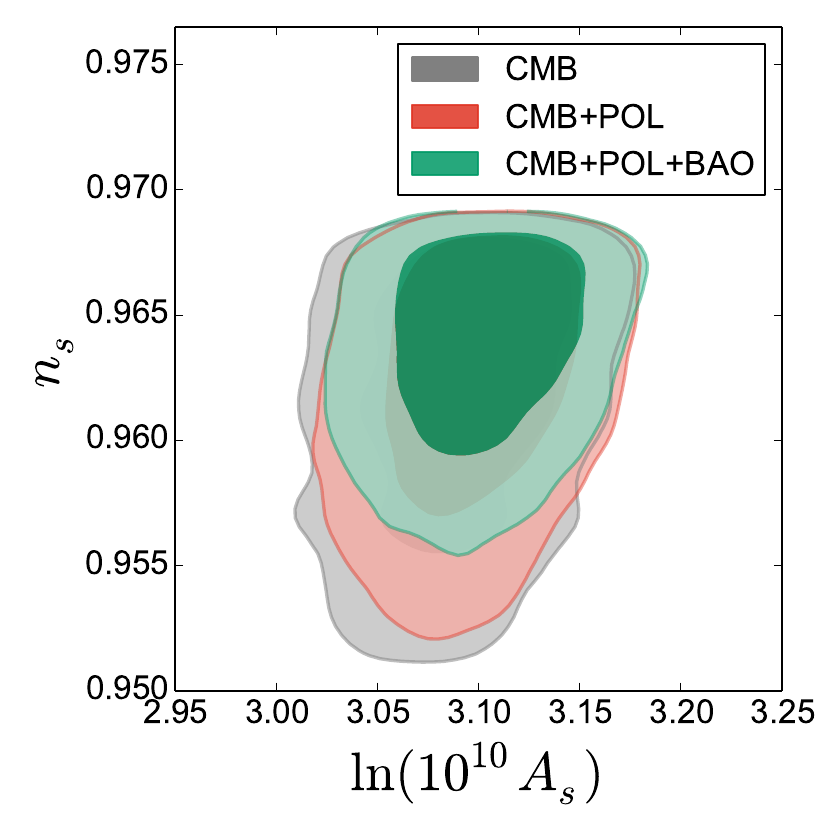} &
\includegraphics[width=.4\textwidth]{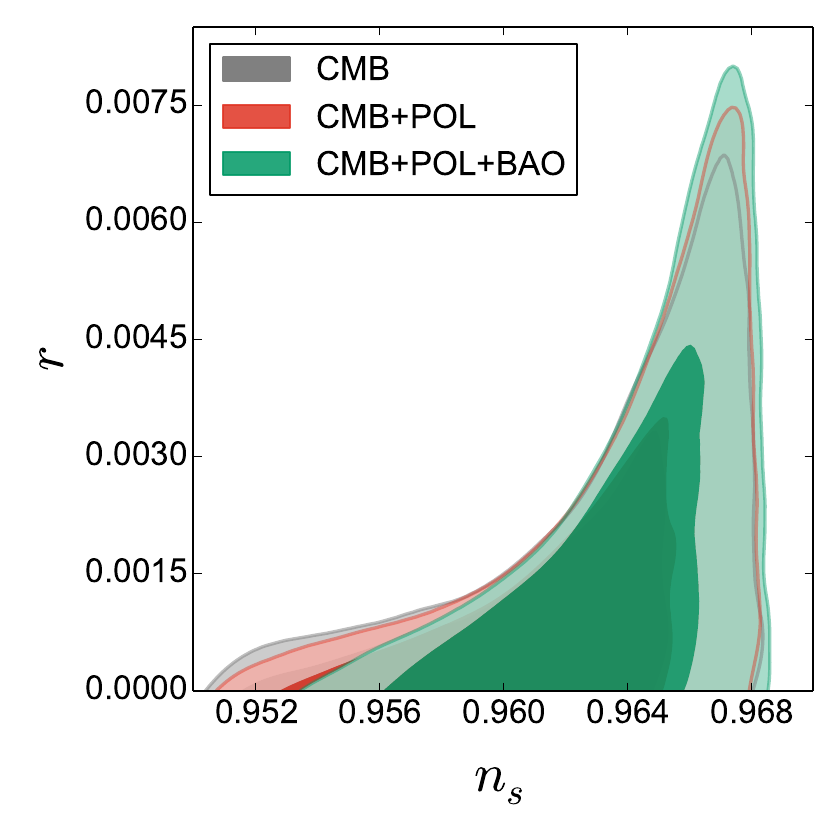}
\end{tabular}
\caption{The upper panel shows the equivalent to the right panel of Fig.~\ref{fig:asnsbis} but translating $\alpha$ into the derived, usual inflationary parameter $n_s$. The lower panel depicts  $68\%$ and $95\%$~CL allowed regions in the ($n_s$, $r$) plane, which here are derived parameters.}
\label{fig:asns}
\end{center}
\end{figure}

The two dimensional allowed contours in the parameters governing the inflationary approach analyzed here, see Eqs.~(\ref{eos}) and (\ref{cc}), are shown in Fig.~\ref{fig:asnsbis}. The first two panels show the $68\%$ and $95\%$~CL allowed contours in the ($\alpha$, $\beta$) and ($\alpha$, $\delta$) planes. The pattern of the first panel agrees with the findings of Ref.~\cite{Barranco:2014ira}. On the other hand, if $\alpha$ decreases, $\delta$ should increase, otherwise the value of the tensor-to scalar ratio will saturate current bounds (see the first of Eqs.~(\ref{derived}). Therefore, there exists a (very) mild negative correlation between $\alpha$ and $\delta$. In Fig.~\ref{fig:asnsbis} we also depict the $68\%$ and $95\%$~CL allowed regions in the ($\ln[10^{10}A_s]$, $\alpha$) plane. Notice that they show a mild anti-correlation, that will be mapped into the usual positive correlation between the $n_s$ and $A_s$ parameters~\cite{Ade:2015xua}, as illustrated in the upper panel of Fig.~\ref{fig:asns}.

Concerning the standard cosmological $\Lambda$CDM parameters $\Omega_{\rm b} h^2$, $\Omega_{\rm c} h^2$, $\theta_{\rm s}$, $\tau$  and $\log[10^{10}A_{s}]$ and the derived quantities $H_0$ and $\sigma_8$, the limits we find here are very similar to those found by the Planck collaboration for the same combination of data sets~\cite{Ade:2015xua} once the additional inflationary parameters, as the scale dependence of the primordial fluctuations (\textit{i.e.}\ the running $\alpha_s$) or the tensor-to-scalar ratio $r$, are also allowed to freely vary. Indeed, the limits we find are slightly more constraining than those quoted by the Planck team analyses for non-zero $\alpha_s$ and/or $r$~\footnote{The errors on the base $\Lambda$CDM parameters are larger when the value of $\alpha_s$ is not set to zero, due to parameter degeneracies.}, as in the phenomenological description of inflation explored here the former two parameters are derived quantities which must follow fixed trajectories for a given number of $e$-folds, \textit{i.e.}, they are not freely varied among some chosen priors. Therefore, our limits on the derived standard inflationary parameters $n_s$ , $r$ and $\alpha_s$ are, in general, much tighter than those quoted by Refs.~\cite{Ade:2015xua,Ade:2015lrj}, as they are further restricted in our inflationary approach, see Eqs.~(\ref{derived}).

\begin{table*}
\begin{center}
\begin{tabular}{lcccc}
\hline \hline
                            & CMB                            & CMB+POL                        & CMB+POL+BAO                    \\  
\hline                                                                                                                                                          
$100\Omega_{\textrm{b}}h^2$ & $2.214\,^{+0.040}_{-0.041}$    & $2.220\,^{+0.029}_{-0.030}$               & $2.225\,^{+0.026}_{-0.027}$               \\
$\Omega_{\textrm{c}}h^2$    & $0.1208\,^{+0.0034}_{-0.0031}$  & $0.1204\,^{+0.0027}_{-0.0023}$  & $0.1197\,^{+0.0020}_{-0.0018}$\\
$100\theta$                 &  $1.0407\,^{+0.0008}_{-0.0009}$            & $1.0407\pm0.0006$                & $1.0408\pm0.0006$                \\
$\tau$                      & $0.077\,^{+0.033}_{-0.034}$    & $0.081\pm0.032$               & $0.083\,^{+0.030}_{-0.031}$               \\
$\ln[10^{10}A_s]$           &  $3.09\,^{+0.06}_{-0.07}$                  & $3.10\pm0.06$                & $3.10\pm0.06$                 \\
$\alpha$                    &  $2.32\,^{+0.46}_{-0.42}$                       & $<2.71$                        &  $2.19\,^{+0.35}_{-0.29}$                       \\
$\beta$                     & {\rm {Unconstrained}}          & {\rm {Unconstrained}}          & {\rm {Unconstrained}}          \\
$\log[\delta]$             & {\rm {Unconstrained}}          & {\rm {Unconstrained}}          & {\rm {Unconstrained}}          \\
\hline
$H_0$                       & $66.8\,^{+1.4}_{-1.5}$               & $67.0\,^{+1.1}_{-1.2}$                  &      $67.3\,^{+0.8}_{-0.9}$               \\
$\sigma_8$                  &  $0.83\pm0.03$                   & $0.83\pm0.03$                  &  $67.3\,^{+0.8}_{-0.9}$        \\
$n_s$                       &  $0.962\,^{+0.007}_{-0.008}$   & $0.962\,^{+0.006}_{-0.007}$    & $0.964\,^{+0.005}_{-0.006}$    \\
$10^3r$                     &$<4.7$                       & $<5.2$                         & $<5.9$                         \\
$10^3\alpha_s$              & $-0.63\,^{+0.11}_{-0.13}$      &$-0.62\,^{+0.10}_{-0.12}$      & $-0.60\,^{+0.08}_{-0.10}$      \\
$c_s$                       & $>0.893$                &  $>0.908$               &  $>0.907$                \\
\hline
\hline
\end{tabular}
\caption{$95\%$~CL constraints on 
the cosmological parameters from the different combinations of data sets explored here.}
\label{tab:results}
\end{center}
\end{table*}

Concerning the tensor-to-scalar ratio $r$, as explicitly show in Ref.~\cite{Barranco:2014ira}, the measured value of the scalar spectral index $n_s\simeq 0.96$ can be related to two different regions of $r$. The first one corresponds to very small values of $r\sim 10^{-3}$, and contains, for instance, Starobinsky models of inflation~\cite{Starobinsky:1980te}, which will be represented in our phenomenological inflationary prescription by $\alpha=2$ and $\beta=1/2$~\cite{Mukhanov:2013tua}. The second region corresponds to much larger values of $r$, as those predicted in the chaotic inflationary scenario. Our numerical analyses showed that current cosmological data isolates the first branch as the allowed one and rejects with very high significance the second one~\cite{Barranco:2014ira}. The limit we find in this work for the tensor-to-scalar ratio $r$ is $r<0.005$ at $95\%$~CL from CMB data. The addition of BAO measurements slightly softens this limit, leading to  $r<0.006$ at $95\%$~CL. These limits are much tighter than those found in Ref.~\cite{Boubekeur:2014xva}, due to \textit{(a)} the \emph{predicted} value of the $r$ parameter is further reduced in the model explored here, which extends that of Ref.~\cite{Boubekeur:2014xva} by including a non-constant sound speed $c_s$; \textit{(b)} the improved CMB measurements used here, which show a strong preference for the low-$r$ region. Figure~\ref{fig:asns}, lower panel, illustrates  the $68\%$ and $95\%$~CL allowed regions in the plane of the ($n_s$, $r$) derived parameters. Finally, the CMB+POL+BAO dataset gives a mean value and $95\%$~CL errors on the running of the spectral index $-0.60\,^{+0.08}_{-0.10}\times 10^{-3}$.

As previously stated, a varying sound speed, $c_s=c_s(\tau)$ can generate non-gaussianties. Assuming the allowed $95\%$ ranges for the $\delta$ parameter quoted above, which modulates the time-dependence of the sound speed during inflation, the maximal non-gaussianity amplitude is $|f^{\text{equil}}_{\text{NL}}|<1$. The current limit on equilateral non-gaussianity from the Planck experiment~\cite{Ade:2015ava} including polarization data is $f^{\text{equil}}_{\text{NL}}=-4\pm 43$, and therefore it does not provide additional constraints in the model explored here. Despite the fact that future surveys, such as the planned SPHEREx~\cite{Dore:2014cca}, based on accurate measurements of both the power spectrum and the bispectrum, could greatly improve the current limits on the amplitude of non-gaussianties of the so-called local type~\cite{dePutter:2014lna}, equilateral non-gaussianity does not give rise to a scale-dependent bias in the matter power spectrum. Higher-order spectra, such as the bispectrum of halos and galaxies, could be enhanced in the presence of equilateral non-gaussianities~\cite{Sefusatti:2007ih}, and future surveys could provide competitive limits to current CMB bounds on $f^{\text{equil}}_{\text{NL}}$~\cite{Hashimoto:2016lmh}, serving as a consistency check of the CMB derived constraints and probing this shape of non-gaussianity at smaller scales. High-redshift, all-sky surveys, reaching $z\sim 5$, could reach $f^{\text{equil}}_{\text{NL}}\sim 1$~\cite{Sefusatti:2007ih,Yokoyama:2013mta}, providing an independent test of the phenomenological inflationary model studied here.

Finally, it is worth computing if there exists a preference in nature for a time-varying inflationary sound speed. For that purpose, we shall compare in the following the $\Delta \chi^2$ for the case $\gamma=1$, $\delta=0$ to the case in which these two parameters are allowed to vary.  Focusing on the CMB+POL data analyses, we find $\Delta \chi^2=1.7$ for $\Delta(dof)=1$, with $dof$ referring to the degrees of freedom involved in each of the parameterizations. The corresponding p-value, which amounts to $0.19$, is not statistically significant and therefore one can conclude that current cosmological data shows no preference for a time-varying sound speed.

\section{Conclusions}
\label{sec:conclusions}

Inflation is currently the most compelling theory to address the standard cosmological problems as well as the origin of the primordial perturbations. In the simplest slow-roll inflationary models, the sound speed of the primordial curvature perturbations is that of the speed of light ($c_s=1$). However, there are plenty of other possible mechanisms equally allowed by current cosmological measurements in which the sound speed is not constant during inflation. In these models, the non-constant sound speed will give rise to scale-dependent features in the primordial power spectrum. Also, a time-varying sound speed will induce an equilateral non-gaussianity amplitude, $f^{\text{equil}}_{\text{NL}}$, which, in general, is expected to be scale-dependent. 

Usually, data analyses are presented in terms of the scalar spectral index $n_s$ and the tensor-to-scalar ratio $r$, which can then be related to a particular model via the inflationary slow-roll parameters. Here we analyze an alternative parametrization due to Mukhanov~\cite{Mukhanov:2013tua,Barranco:2014ira,Boubekeur:2014xva}, in which inflation is described by means of \textit{(a)} an equation of state with two free parameters which can describe a plethora of inflationary models; and \textit{(b)} a sound speed with also two free parameters  which allows us to account as well for non-canonical scenarios. Both the equation of state and the sound speed depend on the number of $e$-folds $N$.  Exploiting the most recent CMB measurements from the Planck satellite, which include temperature and polarization data, and also BAO measurements from a number of galaxy surveys, we have performed Markov Chain Monte Carlo (MCMC) analyses of this phenomenological approach to inflation. 
Despite a time-dependent $c_s$ can arise in theoretically appealing scenarios, current cosmological measurements show no statistical significant preference for such an option.
The obtained bounds on the equation of state parameters (which seem to favour Starobinsky-like scenarios) can be translated into constraints in the usual inflationary parameters $n_s$ and $r$, which are \emph{derived} quantities, rather than freely varying ones. Within this parametrization we are able to discard scale invariance with a significance of about $10\sigma$, and the tensor-to scalar ratio is constrained to be $r<0.005$ at $95\%$~CL. As in the case of the tilt, the running of the spectral index within this phenomenological approach is always negative, and its $95\%$~CL  bounds are $\alpha_s=-0.62\,^{+0.08}_{-0.09} \times 10^{-3}$. These constraints are much tighter than those quoted by the Planck collaboration, as $n_s$, $r$ and $\alpha_s$ are not free parameters in this approach. Given a number of $e$-folds, these quantities are related in a specific way, and must follow fixed trajectories in their respective parameter space~\cite{Barranco:2014ira}. Additional bounds could be placed using the non-gaussianity signals in these models. However, our stringent constraint on the sound speed leads to a very small non-gaussianity amplitude, $|f^{\text{equil}}_{\text{NL}}|<1$, which lies an order of magnitude below current CMB  limits on equilateral non-gaussianities. Future all-sky, high-redshift surveys may provide an additional test of the phenomenological inflationary description provided here, via the measurement of the bispectrum and higher order spectra of biased objects, such as halos or galaxies.
 
\section*{acknowledgments}
The authors thank E.~Giusarma for useful comments on the manuscript. O.M. and H.R. are supported by PROMETEO II/2014/050, by the Spanish
Grant FPA2014--57816-P of the MINECO, by the MINECO Grant
SEV-2014-0398 and by the European Union's Horizon 2020
research and innovation programme under the Marie Sk{\l}odowska-Curie grant
agreements 690575 and 674896.
S.G. is supported by the research Grant {\sl Theoretical Astroparticle Physics} number 2012CPPYP7 
under the Program PRIN 2012 funded by the Ministero dell'Istruzione, Universit\`a e della Ricerca (MIUR).

\bibliography{bibliography}

\end{document}